\documentstyle[12pt,epsfig]{article} 
\begin{document}
\title{Quantum nondemolition measurements of a particle in an inhomogeneous gravitational field.}
\author{ A. Camacho
\thanks{email: acamacho@aip.de} \\
Astrophysikalisches Institut Potsdam. \\
An der Sternwarte 16, D--14482 Potsdam, Germany.}

\date{}
\maketitle

\begin{abstract}
\centerline{\it Dedicated to Heinz Dehnen in honour of his $65^{th}$ birthday.}
\bigskip

In this work we obtain a family of quantum nondemolition variables for the
case of a particle moving in an inhomogeneous gravitational field. Afterwards,
we calculate the corresponding  propagator, and deduce the probabilities
associated with the possible measurement outputs. The comparison, with the case
in which the position is being monitored, will allow us to find the
differences with respect to the case of a quantum demolition measuring process.
\end{abstract}

\newpage
\section{Introduction}
\bigskip

One of the fundamental conceptual difficulties in modern physics bears the name quantum measurement problem,
and comprises the result of a measurement when quantum effects are taken into account.
In the first decades of the XX century the feasibility of analyzing this problem
in the ex\-pe\-ri\-mental realm was an impossible task, but the technology in the 1980s finally allowed the possibility of performing repeated measurements on single quantum systems, in which
some of the most striking features of quantum measurement theory could be tested [1].

In the context of the orthodox quantum theory a measurement is described by von Neumann's reduction postulate,
nevertheless, as we already know, this proposal shows conceptual difficulties [2]. In order to solve these problems some
formalisms, which are e\-qui\-va\-lent to each other, have been proposed [3].
In this direction some works, stemming from the aforementioned proposals, have been done. They
render theoretical predictions that could be tested against the experiment [4, 5, 6].

Neverwithstanding, all these theoretical predictions have been done in the realm
of the so called quantum demolition (QD) measurements, in which an absolute limit on the measurability of the
measured quantity is always present. This limit is a direct consequence of Heisenberg's uncertainty principle [1].
As we already know, there are another kind
of measuring processes, in which this standard quantum limit can be avoided, they are called
quantum nondemolition (QND) measurements [1]. The idea here is to measure a variable such that the unavoidable disturbance
of the complementary one does not disturb the evolution of the chosen variable, this idea was pioneered by Braginsky, Vorontsov, and Khalili [7].

The relevance of the understanding of the measurement problem has not only academic importance,
it possesses also practical importance, for instance, it plays a relevant role in the comprehension of the measurement of the position of the elements of a gravitational--wave antenna [1].
Another point in connection with the quantum measurement problem is its possible relation with
gravity, namely it could also shed some light upon another fundamental problem in modern physics, 
i.e., the question around the validity at quantum level of the equivalence principle [8, 9].

In this work we find a family of QND variables for the case of a particle immersed in an inhomogeneous gravitational field.
Afterwards, its propagator, when these QND variables are measured, is calculated,
and the probabilities, corresponding to different measurement outputs, are deduced.
Finally, we compare this case with the results of a quantum demolition situation [6],
namely to the case in which the position of our quantum particle is continuously monitored.
\bigskip

\section{Quantum nondemolition variables}
\bigskip
 
Consider the Lagrangian of a particle, with mass $m$, moving in the Earth's
gravitational field, and located a distance $r$ from the center of it

\bigskip

\begin{equation}
L = {\vec{p}~^2\over 2m} + {GMm\over r} + ct.
\end{equation}
\bigskip

The corresponding Hamiltonian reads

\begin{equation}
H = {\vec{p}~^2\over 2m} - {GMm\over r} + ct.
\end{equation}
\bigskip

At this point we introduce an approximation, namely the particle is located
at a distance $l$ above the Earth's surface, where the condition $l<<R$ is fulfilled, here $R$ denotes the radius of the Earth.
The intention with this approximation is to obtain a path integral, and to be able
to calculate it analytically, even if we consider the effects of a measuring process.
Mathematically this simplification is important, since as is already known [10], even the path integral (here path integral means Feynman's original time--sliced formula) of a quantum particle moving in a Coulombian
potential (without the inclusion of a measuring process) shows mathematical problems, for
instance, the so called path collapse. This aforementioned simplification allows us
to avoid this problem, and also to find in a, more or less, simple way the analytical expression for the corresponding propagator.
Therefore, in a first approach to this problem we consider the case in which the
particle remains near the Earth's surface.

Under these conditions the Hamiltonian becomes

\begin{equation}
H = {\vec{p}~^2\over 2m} + mgl + {m\omega^2\over 2}l^2,
\end{equation}
\bigskip

\noindent where $g = {GM\over R^2}$, $\omega= i\sqrt{{2g\over R}}$, and the constant, that appears in (2), has been chosen equal to $mgR$.
\bigskip

Let us now introduce the operator

\begin{equation}
A = \rho l + \sigma p,
\end{equation}
\bigskip

\noindent where $\rho, \sigma: \Re\rightarrow\Re$.
\bigskip

The differential equation that determines if this last operator defines a QND va\-ria\-ble reads [11]

\begin{equation}
{d\over dt}({\rho\over\sigma})= {1\over m}({\rho\over\sigma})^2 - 2{mg\over R}.
\end{equation}
\bigskip

Defining $f(t) = {\rho\over\sigma}$, we find that expression (5) becomes

\begin{equation}
{d\over dt}(f(t))= {1\over m}(f(t))^2 - 2{mg\over R}.
\end{equation}
\bigskip

Its solution reads 

\begin{equation}
f(t) = - m\sqrt{{2g\over R}}\tanh\Bigl({\sqrt{{2g\over R}}(t - \tau')}\Bigr),
\end{equation}
\bigskip

\noindent here $\tau'$ is a constant.
\bigskip

Hence

\begin{equation}
A(t) = \sigma(t)\Bigg[p - m\sqrt{{2g\over R}}\tanh\Bigl({\sqrt{{2g\over R}}(t - \tau')}\Bigr)l\Bigg].
\end{equation}
\bigskip

The solution given by expression (8) defines a family of functions, and each element of it
is a QND variable. Indeed, any choice for our function $\sigma(t)$ renders a QND variable,
and this function is not determined by the differential equation and remains as a free parameter in
our model.
\bigskip

\section{Propagators and Probabilities}
\bigskip

Let us now introduce a measuring process, namely the observable related to $A(t)$
will be measured continuously. In order to analyze the propagator of our system we will resort to the
so called restricted path integral formalism, in which the measuring process is taken into account
through a weight functional in the path integral [11].
This means that now we must consider
a weight functional, which has to take into account the  effect of the measuring device upon the measured system.
Mathematically, this weight functional restricts the integration domain, i.e., the integration
is performed only over those paths that match with the measurement output.

The problem at this point is to choose a particular expression for this weight functional.
Here our choice is a gaussian form. A justification for this election
stems from the fact that the results coming from a Heaveside weight functional [12]
and those coming from a gaussian one [13] coincide up to the order of magnitude.
At this point it is noteworthy to add that even the possibility of having measuring processes with this kind of functionals
has already been discussed [14].

This kind of weight functional has been employed to analyze the response of a
gravitational wave of Weber type [11], the measuring process of a gravitational
wave in a laser--interferometer [15], or even to explain the emergence of the classical
concept of time [16].

Therefore, in a first approach to this topic we may consider a measuring device whose weight functional has a gaussain form.

Hence under these conditions the propagator becomes

{\setlength\arraycolsep{2pt}\begin{eqnarray}
U_{[a(t)]}(z'', z') = \int_{z'}^{z''}d[l]d[p]\exp{\Big[}{i\over\hbar}\int_{\tau '}^{\tau ''}({\vec{p}~^2\over 2m} - mgl - {m\omega^2\over 2}l^2)dt{\Big]} \times\nonumber\\
\exp{\Big(}  -{1\over T\Delta a^2}\int _{\tau '}^{\tau ''}[A(t) - a(t)]^2dt{\Big)},
\end{eqnarray}}
\bigskip

\noindent where $\Delta a^2$ is the experimental error associated with the
measuring of $A$, and $T = \tau'' - \tau'$ denotes the time that the experiment lasts.
\bigskip

Expression (9) may be rewritten as follows

{\setlength\arraycolsep{2pt}\begin{eqnarray}
U_{[a(t)]}(z'', z') = \exp\Bigl[-{1\over T\Delta a^2}\int _{\tau '}^{\tau ''}a(t)^2dt\Bigr]\times\nonumber\\
\int_{z'}^{z''}d[l]d[p]\exp{\Bigg(}{i\over\hbar}\int_{\tau '}^{\tau ''}{\Bigg[}({1\over 2m} + {i\hbar\over T\Delta a^2}\sigma^2)\vec{p}~^2 \nonumber\\
- 2{i\hbar\sigma\over T\Delta a^2}\Bigl(a + m\sigma\sqrt{{2g\over R}}\tanh[\sqrt{{2g\over R}}(t - \tau')]l\Bigr)p{\Bigg]}dt{\Bigg)}\times\nonumber\\
\exp{\Bigg(}{i\over\hbar}\int_{\tau '}^{\tau ''}{\Bigg[}\Bigl(2{i\hbar m^2\sigma^2g\over TR\Delta a^2}\tanh^2[\sqrt{{2g\over R}}(t - \tau')] - {m\omega^2\over 2}\Bigr)l~^2\nonumber\\
+\Bigl(2{i\hbar am\sigma\over T\Delta a^2}\sqrt{{2g\over R}}\tanh[\sqrt{{2g\over R}}(t - \tau')] - mg\Bigr)l{\Bigg]}dt{\Bigg)}.
\end{eqnarray}}
\bigskip

The path integral on the momenta is readily calculated [17]

{\setlength\arraycolsep{2pt}\begin{eqnarray}
\int_{z'}^{z''}d[p]\exp{\Bigg(}{i\over\hbar}\int_{\tau '}^{\tau ''}{\Bigg[}({1\over 2m} + {i\hbar\over T\Delta a^2}\sigma^2)\vec{p}~^2 \nonumber\\
- 2{i\hbar\sigma\over T\Delta a^2}\Bigl(a + m\sigma\sqrt{{2g\over R}}\tanh[\sqrt{{2g\over R}}(t - \tau')]l\Bigr)p{\Bigg]}dt{\Bigg)}=\nonumber\\
\exp{\Bigg(}{4i\hbar m\over T\Delta a^2(T\Delta a^2 + 2im\hbar)}\int_{\tau '}^{\tau ''}{\Bigg[}a^2 + m^2\sigma^2{2g\over R}\tanh^2[\sqrt{{2g\over R}}(t - \tau')]l^2\nonumber\\
+ 2ma\sigma\sqrt{{2g\over R}}\tanh[\sqrt{{2g\over R}}(t - \tau')]l{\Bigg]}{\Bigg)}.
\end{eqnarray}}
\bigskip

This last result allows us to rewrite the propagator as follows

{\setlength\arraycolsep{2pt}\begin{eqnarray}
U_{[a(t)]}(z'', z')= \exp{\Bigg(}-{1\over T\Delta a^2}{(T\Delta a^2 - 2im\hbar)^2\over (T\Delta a^2)^2 + (2m\hbar)^2}\int _{\tau '}^{\tau ''}a(t)^2dt{\Bigg)}\times\nonumber\\
\int_{z'}^{z''}d[l]\exp{\Bigg(}{i\over\hbar}\int_{\tau '}^{\tau ''}{\Bigg[}\Bigl(-2{m^2\sigma^2g\over TR\Delta a^2}\tanh^2[\sqrt{{2g\over R}}(t - \tau')] - {im\omega^2\over 2\hbar}\nonumber\\
+ {8i\hbar m^3g\sigma^2\over TR\Delta a^2(T\Delta a^2 + 2im\hbar)}\tanh^2[\sqrt{{2g\over R}}(t - \tau')]\Bigr)l^2\nonumber\\
+\Bigl(-2{am\sigma\over T\Delta a^2}\sqrt{{2g\over R}}\tanh[\sqrt{{2g\over R}}(t - \tau')] - {img\over\hbar} \nonumber\\
+ {8i\hbar m^2a\sigma\over T\Delta a^2(T\Delta a^2 + 2im\hbar)}\sqrt{{2g\over R}}\tanh[\sqrt{{2g\over R}}(t - \tau')]\Bigr)l{\Bigg]}{\Bigg)}.
\end{eqnarray}}
\bigskip

Once again we find a gaussian path integral, which can be easily calculated

{\setlength\arraycolsep{2pt}\begin{eqnarray}
U_{[a(t)]} = \exp{\Bigg(}\int _{\tau '}^{\tau ''}{\Bigg[}-\alpha a^2(t) +
{{4a^2(t)gm^2\alpha^2\over R} - {m^2g^2\beta^2\over 2\hbar^2} +
{2ia(t)m^2\alpha\beta g\sqrt{{2g\over R}}\over\hbar}\over {4m^2g\alpha\over R} + i{m\omega^2\beta^2\over \hbar}}{\Bigg]}dt{\Bigg)}.
\end{eqnarray}}
\bigskip

In this last expression, in order to simplify it, we have introduced two definitions, namely,
$\alpha = {(T\Delta a^2)^2 - 4m^2\hbar^2 - 4im\hbar T\Delta a^2\over
T\Delta a^2[(T\Delta a^2)^2 + 4m^2\hbar^2]}$ and  $\beta(t) = {\coth[\sqrt{{2g\over R}}(t - \tau')]\over \sigma(t)}$.
\bigskip

Employing (13) we may calculate the probability (it is related to the total probability of $a(t)$), $P_{a(t)} = \vert U_{[a(t)]}(z'', z')\vert^2$, of having as
measurement output, for our QND variable $A$, the function $a(t)$

{\setlength\arraycolsep{2pt}\begin{eqnarray}
P_{[a(t)]}(z'', z') = \exp{\Bigg(}\int _{\tau '}^{\tau ''}{\Bigg[}-2{(T\Delta a^2)^2 - 4m^2\hbar^2\over T\Delta a^2[(T\Delta a^2)^2 + 4m^2\hbar^2]}a^2\nonumber\\
+ {8a^2m^2g\over R(\gamma^2 + \Gamma^2)}
{\gamma[(T\Delta a^2)^2 - 4m^2\hbar^2]- 4\Gamma m\hbar T\Delta a^2\over T\Delta a^2[(T\Delta a^2)^2 + 4m^2\hbar^2]}
+ {4am^2\beta g\sqrt{{2g\over R}}\Gamma\over\hbar(\gamma^2 + \Gamma^2)}\nonumber\\
-{m^2g^2\beta^2\over\hbar^2}
{{4m^2g\over TR\Delta a^2}{(T\Delta a^2)^2 - 4m^2\hbar^2]\over (T\Delta a^2)^2 + 4m^2\hbar^2]}
\over[{4m^2g\over TR\Delta a^2}{(T\Delta a^2)^2 - 4m^2\hbar^2]\over (T\Delta a^2)^2 + 4m^2\hbar^2]}]^2 +[{m\omega^2\beta^2\over\hbar} - {16m^3g\hbar\over R[(T\Delta a^2)^2 + 4m^2\hbar^2]}]^2}{\Bigg]}dt{\Bigg)}.
\end{eqnarray}}
\bigskip

Here $\gamma = {4m^2g\over R} - {4m^2\omega^2\beta^2(T\Delta a^2)^2\over (T\Delta a^2)^2 + (2m\hbar)^2}$
and $\Gamma = {m\omega^2\beta^2T\Delta a^2\over\hbar}{(T\Delta a^2)^2 - (2m\hbar)^2\over (T\Delta a^2)^2 + (2m\hbar)^2}$.

\bigskip

\section{Conclusions}
\bigskip

We have found a family of QND variables for the case of a particle immersed in an inhomogeneous
gravitational field, and afterwards, we have calculated the co\-rres\-ponding propagator and probabilities, when
these QND variables are subject to a continuous quantum measurement.

Expression (14) gives us the probability of having as measurement output, for our observable $A(t)$,
the function $a(t)$. In connection with (14) there is an interesting case, namely, if
$T\Delta a^2 = 2m\hbar$, then we obtain that $P_{[a(t)]} = 1$, in other words, all
the possible measurement outputs have the same probability. This is a peculiar situation,
indeed, if we remember how the effects of the measuring device have been introduced,
we may notice that we have considered a weight functional, which, as has already been mentioned,
restricts the integration domain [11], i.e., only those paths matching with the measurement
output are taken into account in the path integral.
Nevertheless, if the resolution of the measuring device satisfies the condition
$\Delta a^2 = {2m\hbar\over T}$, then all paths have the same probability.
In this sense this case is as if there were no measuring process. Indeed, in the
situation in which there is no measurement, all the paths have the same probability [18].
This quantum feature appears only in connection with QND measurements, clearly, in the case
of QD measurements the effects of the weight functional do not disappear in the
expression for the probability [6]. As we already know, a QND measurement could be classified
as a classical measuring process, in the sense that there is no standard quantum limit [11].
But as it has already been shown, there are QND cases in which all the possible trajectories
have the same probability, a situation which does not match with a classical case,
in which only one trajectory has non--vanishing probability. This feature should be 
no surprise, if we consider a one--dimensional harmonic oscillator (see page 106, equation (6.32)
of [11]), then considering the limit of zero experimental resolution we obtain, once again, this behaviour.
The contribution of the present work comprises the statement that this characteristic appears, 
not only if the resolution vanishes, but also when it satisfies a certain relation, the one involves the
duration of the measuring process, and also the mass of the particle.

Looking at (14) we may notice that the mass of the test particle, $m$, appears once again in the probability,
i.e., gravity is at quantum level not a purely geometric effect [19]. Another point
comprises the fact that this mass appears but not always in the relation $m/\hbar$, as happens in
the Collela, Werner, and Overhauser experiment [20], or in the case of a QD experiment [6].

Let us now consider the limit $\Delta a^2 \rightarrow 0$. If we do this, we find that
\bigskip

{\setlength\arraycolsep{2pt}\begin{eqnarray}
P_{[a(t)]}(z'', z') \rightarrow \exp\Bigl[{4m^2GM\over R\beta^4\Delta a^2}\Bigr].
\end{eqnarray}}
\bigskip

Clearly, we may notice that
there is no standard quantum limit, in other words, we may measure $A$ with an arbitrarily small error,
and in consequence all the necessary information can be extracted.
Another feature that we may notice from expression (15) involves the fact that
this probability does not depend upon the measurement output, i.e., upon $a(t)$.

We have two experimental preparations in which all the trayectories have the same probability, i.e.,
if $\Delta a^2 = {2m\hbar\over T}$, or if $\Delta a^2 \rightarrow 0$.
In other words, if we know that in a certain experiment all the trayectories have
the same probability, it could not be determined, without further infomation, if the experiment was carried
out with a vanishing experimental error, or with $\Delta a^2 = 2m\hbar/T$.

We could have a very small experimental error, but the case $\Delta a^2 \rightarrow 0$ 
is an idealization, and this limit has to be understood in the sense that if the 
experimental resolution is much smaller that all the relevant physical variables, then 
we could expect to have a probability independent of the measurement outputs. Experimentally 
this case could be a very difficult one, consider, for example, the current experimental
resolution in the case of Paul or Penning traps [21], which lies very far from this idealization.

At this point it is noteworthy to comment what else can be learned from the case
$\Delta a^2 \rightarrow 0$. It is readily seen from expression (15) that this  
expression contains information about the potential, namely, about the coupling
constant (Newton's constant), $G$, the source of the gravitational field, $M$,
and also about the geometry of this source, $R$. In other words, if we had, instead
of a gravitational field, an electric field, then we would have an expression containing
the source of the electric field, of the coupling constant between the involved charges,
and also about the radius of the source of this field.
 
An additional interesting characteristic of expression (15) comprises the fact that it does not
include Planck's constant, $\hbar$, but it does include the mass of the test particle, $m$. Here we have a
radical difference with respect to the situation in which we measure a QD variable [6, 19, 20].
Indeed, in the QD measurement case, mass and  Planck's constant appear always as a function of $m/\hbar$.

Regarding this last remark we must add that in order to see nontrivial quantum mechanical effects of gravity
it is not neccesary, as was believed [19], to study effects in which $\hbar$ appears
explicitly. At most we could assert that this is necessary in the case of QD experiments, but
not if the corresponding measurements are carried out in the realm of QND proposals.

At this point a word must be said concerning the operational meaning of the concept of continuous quantum measurement. 
As we know, measurements can not be considered instantaneous (they have a finite duration), and this characteristic 
endows the idea of continuous measurement with an important physical meaning. A more elaborated model 
comprises a continuous measurment as a series of instantaneous measurements with the interval between them tending to zero [22].
Nevertheless, this is a controversial issue, but the work in the direction of Paul and
Penning traps [21] could be, in a near future, a way to perform this kind of experiments [23]. 

To finish, let us comment an additional characteristic of expression (14). We know that there
are some QND measurements in which an absolute limit may appear. For instance,
if the linear momentum of a free particle is monitored, this aforementioned limit may emerge,
when the instantaneous measurement of position of the test particle, before and after the monitoring
of the linear momentum, is done (see page 99 of reference [11]). Clearly, position and linear momentum
are canonical conjugate variables to each other, and that is why this absolute limit appears. At this
point we may wonder why (14) has no absolute limit, in our case, position, before and after monitoring of $A(t)$, has also been instantaneously measured, i.e.,
$z'$ and $z''$ are present from the very begining in our mathematical expressions, see (9).

The answer to this question stems from the fact that in our case the measured quantity, $A(t) = \sigma(t)\Bigl[p - m\sqrt{{2g\over R}}\tanh\Bigl({\sqrt{{2g\over R}}(t - \tau')}\Bigr)l\Bigr]$,
is not the canonical conjugate variable of position, $l$. We may understand better this point
noting that in the present case we deal with a one--dimensional harmonic oscillator (it has a complex frequency but
this feature plays in this discussion not role at all) subject to a measuring process. In the case of a one--dimensional
harmonic oscillator, the canonical conjugate variable of $p(t) = -m\omega l(0)\sin{(\omega t)} + p(0)\cos{(\omega t)}$,
is $l(t) = l(0)\cos{(\omega t)} + {p(0)\over m\omega}\sin{(\omega t)}$, and not $l(t)$ [19].
If we choose $\sigma(t) = \cosh\Bigl({\sqrt{{2g\over R}}(t - \tau')}\Bigr)$,
then we obtain a very si\-mi\-lar situation to a one--dimensional harmonic oscillator case. Of course,
trigonometric functions have to be substituted with the corresponding hyperbolic ones,
but this change is only a consequence of the fact that in the case of a particle moving in
an inhomogeneous gravitational field (with the approximation done after (2)) the frequency of
the emerging harmonic oscillator is complex.

These last remarks explain why expression (14) contains no absolute limit, it can emerge only if the canonical
conjugate variable associated to $A(t)$ were measured, before and after the  monitoring of $A(t)$.
To make this statement more precise, let us choose $\sigma(t) = \cosh\Bigl({\sqrt{{2g\over R}}(t - \tau')}\Bigr)$, therefore an
absolute limit could appear in our case if the variable $Q(t) = {p\over m\sqrt{{2g\over R}}}\sinh\Bigl({\sqrt{{2g\over R}}(t - \tau')}\Bigr) +~\cosh\Bigl({\sqrt{{2g\over R}}(t - \tau')}\Bigr)l$
were measured, instantaneously, before and after the monitoring of $A(t)$.
At this point we must underline that the appearance of this absolute limit does not mean that $A(t)$ is a QD variable [11].
\bigskip

\Large{\bf Acknowledgments}\normalsize
\bigskip

The author would like to thank A. A. Cuevas--Sosa and  A. Camacho--Galv\'an for their
help, and D.-E. Liebscher for the fruitful discussions on the subject. 
The hospitality of the Astrophy\-si\-ka\-li\-sches Institut Potsdam is also kindly acknowledged. 
This work was supported by CONACYT (M\'exico) Posdoctoral Grant No. 983023.

\end{document}